# Hyperons in Neutron Stars across the Observed Mass Range: Insights from ΛN and ΛΛ Interactions within a Microscopic Framework

A. Mohammad Ali Looee[1], M. Shahrbaf[2], and H. R. Moshfegh[1,3]
[1] Department of Physics, University of Tehran, North Karegar Avenue, Tehran 14395-547, Iran
[2] Institute of Theoretical Physics, University of Wroclaw, Max Born pl. 9, 50-204 Wroclaw, Poland
[3] Centro Brasileiro de Pesquisas Fısicas, Rua Dr. Xavier Sigaud, 150, URCA, Rio de Janeiro CEP 22290-180, RJ, Brazil

## Abstract

We investigate the equation of state (EOS) and macroscopic properties of neutron stars (NSs) and hyperonic stars within the framework of the lowest-order constrained variational (LOCV) method, extended to include interacting $\Lambda$ hyperons. The nucleon–nucleon interaction is modeled using the AV18 potential supplemented by Urbana three-body forces, while $\Lambda N$ and $\Lambda\Lambda$ interactions are described by phenomenological spin- and parity-dependent potentials calibrated to available hypernuclear data. Cold, charge-neutral, and $\beta$-equilibrated matter composed of neutrons, protons, electrons, muons, and $\Lambda$ hyperons is considered. We compute particle fractions, chemical potentials, the EOS, speed of sound, tidal deformability, and stellar structure by solving the Tolman–Oppenheimer–Volkoff equations, and compare our results with recent NICER and gravitational-wave observations. The inclusion of $\Lambda$ hyperons leads to EOS softening, reducing the maximum NS mass from $2.34\,M_\odot$ to $2.07\,M_\odot$, while keeping it consistent with the $2\,M_\odot$ mass constraint. At $1.4\,M_\odot$, the model satisfies observational limits on radius and tidal deformability, with the $\Lambda$ onset occurring below this mass. Comparison with other microscopic and relativistic mean-field models shows that our EOS remains consistent with the allowed pressure–energy density range, while also permitting even canonical-mass NSs of about $1.4\,M_\odot$ to accommodate hyperons. These results suggest that hyperons can appear in NSs across the observed mass range without violating current astrophysical constraints, and that the extended LOCV method provides a consistent, microscopic approach to modeling dense hypernuclear matter.

*Unified Astronomy Thesaurus concepts:* Neutron stars (1108); Neutron star cores (1107)

## 1. Introduction

Neutron stars (NSs) are among the densest known objects in the universe, serving as natural laboratories for studying matter under extreme physical conditions.

One of the main challenges in nuclear astrophysics is determining the equation of state (EOS) that governs the behavior of nuclear matter inside NSs, especially in their cores, where densities can reach several times the nuclear saturation density (J. M. Lattimer & M. Prakash 2001; P. Haensel et al. 2007; J. Schaffner-Bielich 2020; G. F. Burgio et al. 2021; H. Tong et al. 2022; S. Elhatisari et al. 2024). At such extreme densities, the chemical potentials of particles rise significantly and may reach the rest masses of heavier baryons. This opens the possibility for the emergence of new degrees of freedom beyond neutrons and protons. In recent years, the possible appearance of hyperons in dense baryonic matter has received considerable attention in both phenomenological and microscopic studies (I. Vidaña 2018; R. Karimi & H. R. Moshfegh 2023; F. M. Jangal et al. 2025; H. Tong et al. 2025a). The impact of hyperons has also been investigated in the context of binary NS mergers (H. Kochankovski et al. 2025). Simulations suggest that hyperons increase the dominant postmerger gravitational-wave frequency by several percent, potentially offering an observable signature of exotic degrees of freedom. Hyperonic EOSs also tend to produce slightly enhanced dynamical mass ejection, which could influence kilonova signals and serve as an indirect probe of hyperon formation. More broadly, macroscopic stellar properties may encode the onset of hyperons. For example, a positive slope in the mass–radius relation near $1.4\,M_\odot$ (M. Ferreira & C. Providência 2025) or a strongly negative curvature in $R(M)$ (A. Bauswein et al. 2025) have been proposed as indicators of hyperonic cores. Bayesian studies using simulated observational data further suggest that next-generation X-ray telescopes, such as eXTP and STROBE-X, could distinguish between nucleonic and hyperonic EOSs with high confidence (C. Huang et al. 2025).

Experimental studies of hypernuclei have tightly constrained the $\Lambda N$ interaction, which is well established as attractive, with a single-particle potential depth of $U_\Lambda \approx -28$ MeV at saturation density (E. Friedman & A. Gal 2023, 2025a). In contrast, $\Sigma$ hyperons experience a predominantly repulsive potential, $U_\Sigma(n_0) \approx 30 \pm 20$ MeV, preventing bound $\Sigma$ states (L. Tolos & L. Fabbietti 2020; J. Haidenbauer et al. 2023; A. Jinno et al. 2025). Evidence for $\Xi$ hypernuclei remains scarce, but current estimates suggest a weakly attractive potential, $U_\Xi \approx -14$ MeV, though uncertainties remain due to possible repulsive three-body $\Xi NN$ forces (E. Friedman & A. Gal 2022, 2025b). A broad range of studies, particularly within relativistic mean-field (RMF) theory, have investigated hyperons in the neutron star EOS, typically calibrating hyperon–meson couplings to empirical single-particle potentials at saturation density (S. Weissenborn et al. 2013; M. Fortin et al. 2016; F. M. Jangal et al. 2025). These models generally predict $\Lambda$, $\Sigma$, and $\Xi$ hyperon onset at (2–3)$\rho_0$ (nuclear saturation density). Beyond RMF, microscopic many-body approaches-such as Brueckner–Hartree–Fock (BHF; Z. Li & H.-J. Schulze 2008; I. Vidana et al. 2011; J. J. Li et al. 2018), and quantum Monte Carlo methods (S. Gandolfi 2013)—employ realistic hyperon–nucleon and hyperon–hyperon interactions, often derived from Nijmegen models. The auxiliary field diffusion Monte Carlo approach is based on simplified phenomenological interactions (D. Lonardoni et al. 2015). Recent BHF studies incorporating chiral $\Lambda N$ interactions constrained by ALICE femtoscopic data, together with lattice-QCD $\Lambda\Lambda$ and $\Xi N$ potentials from the Hadrons to Atomic nuclei from Lattice (HAL) QCD Collaboration, predict a low maximum mass of 1.3–1.4 $M_\odot$ (I. Vidana et al. 2025). Chiral effective field theory (EFT) has also been extended to hypernuclear matter up to $\mathrm{N^2LO}$ in the nonrelativistic sector, while relativistic versions remain at LO due to limited experimental constraints (J.-H. Chen et al. 2025). Furthermore, nuclear lattice EFT (NLEFT) combines EFT with a Euclidean spacetime lattice and Monte Carlo methods. By mapping spacetime to a finite grid on which the nuclear interactions are formulated, it enables ab initio calculations that now simultaneously describe single- and double-hypernuclei (light to medium mass), nuclear and hypernuclear matter EOS, and NS properties (H. Tong et al. 2025a, 2025b, 2025c). Complementary information from double-strangeness hypernuclei, observed in High Energy Accelerator Research Organization (KEK) and Japan Proton Accelerator Research Complex (J-PARC) emulsion experiments, continues to provide valuable constraints on $\Lambda\Lambda$ and $\Xi N$ interactions, which will improve the microscopic approaches (K. Miwa et al. 2025).

While most theoretical approaches agree that hyperons—particularly the $\Lambda$—are likely to appear in NSs cores, their inclusion generally softens the EOS, lowering the maximum mass below the $\sim 2\,M_\odot$ constraint set by heavy pulsars such as PSR J0348+0432 (J. Antoniadis et al. 2013) and PSR J0740+6620 (E. Fonseca et al. 2021). This long-standing tension, known as the hyperon puzzle (I. Bombaci 2017), has motivated numerous proposed solutions, including a phase transition to deconfined quark matter (G. Baym et al. 2018; M. Shahrbaf et al. 2020a, 2020b), modified gravity (A. V. Astashenok et al. 2014), and repulsive hyperon three-body forces (TBFs; I. Vidana et al. 2011; Y. Yamamoto et al. 2014). More recent ab initio studies using NLEFT, with both two- and three-body hyperon interactions, have sought to reduce the softening and yield EOSs consistent with astrophysical limits (H. Tong et al. 2025c).

Alternative scenarios include quarkyonic matter models, where quark substructure delays hyperon onset to higher densities, suppressing their low-density appearance (Y. Fujimoto et al. 2024). The role of the symmetry energy at high density has also been highlighted: hyperonic matter can exhibit a pronounced peak in the sound speed at (3–4)$\rho_0$, compatible with multimessenger constraints (J.-T. Ye et al. 2025). Overall, these studies suggest that suppressing the $\Lambda$ fraction through repulsive momentum-dependent or three-body $\Lambda NN$ interactions is essential for obtaining a viable EOS and resolving the hyperon puzzle in the inner core.

Building on the challenges posed by the hyperon puzzle and the need for EOS models with minimal phenomenological assumptions, we employ a fully microscopic, potential-based approach, the lowest-order constrained variational (LOCV) method, to extract the EOS of dense baryonic matter. The LOCV method has been successfully applied to pure neutron matter, symmetric and asymmetric nuclear matter, and $\beta$-stable matter at both zero and finite temperature (H. Moshfegh & M. Modarres 1998; M. Modarres & H. Moshfegh 2000, 2004; M. Modarres & A. Tafrihi 2023). It offers a parameter-free framework grounded in realistic two-body correlations and thermodynamically consistent variational principles. Recently, we extended the LOCV formalism to include strange degrees of freedom by incorporating interacting $\Lambda$ hyperons (M. Shahrbaf et al. 2019; M. Shahrbaf & H. R. Moshfegh 2019). For the $\Lambda$ interactions, phenomenological spin-parity-dependent potentials based on the SU(3) flavor symmetry of meson–baryon coupling constants have been employed (E. Hiyama et al. 2002, 2006; H. Togashi et al. 2016) to obtain the EOS of symmetric and asymmetric hypernuclear matter.

Given the limited and uncertain experimental data on $\Sigma$ and $\Xi$ hyperons, along with the repulsive nature of the $\Sigma$-nucleus potential (J. R. Stone et al. 2022) and the relatively large mass of the $\Xi$, it is plausible that their appearance in the core of NSs may occur only at very high densities. Consequently, the present study restricts its analysis to the inclusion of the $\Lambda$ hyperon, which is expected to appear at comparatively lower densities and thus plays a more significant role in the EOS and NS structure.

Nevertheless, there are several approaches that incorporate these heavier hyperons into the modeling of NS matter (K. A. Maslov et al. 2015; A. Sedrakian et al. 2023; S.-Y. Ding et al. 2025). Some of these studies extend the analysis to hot and dense hypernuclear matter equations of state, which are particularly relevant for numerical simulations of proto-NSs, core-collapse supernovae, and binary NS mergers (J. R. Stone et al. 2021; H. Kochankovski et al. 2023; S. Tsiopelas et al. 2024). While certain models successfully produce hyperonic stars with masses reaching the observed $2\,M_\odot$ threshold, not all of them are consistent with the tidal deformability constraints derived from the GW170817 event (B. P. Abbott et al. 2018).

In the current work, we apply this extended LOCV method using the AV18 potential (R. B. Wiringa et al. 1995) supplemented by Urbana TBF (S. Goudarzi & H. R. Moshfegh 2015; S. Goudarzi & H. Moshfegh 2019) for nucleon–nucleon, and updated hyperon–hyperon and hyperon–nucleon interactions to compute the composition of dense matter.

The remainder of this paper is structured as follows: Section 2 and its subsections are devoted to a detailed description of the LOCV method and the relevant features of nucleon–nucleon and hyperon–baryon interactions. In Section 3, we present the EOS and composition of the $\beta$-stable hypernuclear matter. Section 4 discusses the corresponding stellar structure. Finally, Section 5 summarizes our main findings and conclusions.

## 2. Formalism

### 2.1. LOCV Method

To describe the EOS of dense baryonic matter, we adopt the LOCV (M. Modarres & A. Tafrihi 2023) method, a microscopic variational approach based on minimizing the total energy functional of the system under a normalization constraint. The energy functional per particle is given by:

$$E[f] = \frac{1}{A}\frac{\langle\Psi|H|\Psi\rangle}{\langle\Psi|\Psi\rangle} = E_1 + E_2 + ..., \quad (1)$$

where $\Psi(1\ldots A) = F(1\ldots A)\varphi(1\ldots\mathrm{A})$ is the correlated many-body wave function, with $\varphi$ the Slater determinant of noninteracting single-particle states and $F$ the correlation operator that incorporates interparticle interactions. Under the Jastrow approximation, the correlation operator $F$ is written as a product of pair correlations:

$$F = S\prod_{i<j} f(ij), \tag{2}$$

where $S$ is the symmetrization operator. The Hamiltonian is written as:

$$H = \sum_{i=1}^{A}\frac{p_i^2}{2m_i} + \sum_{i<j} v(ij), \tag{3}$$

where $v(ij)$ denotes the interaction two-body potential and $m_i$ is the mass of the $i$th baryon.

The term of a one-body cluster energy, $E_1$, represents the kinetic energy of an ideal Fermi gas:

$$E_1 = \sum_i \frac{3\hbar^2 k_{F_i}^2}{10 m_i}, \tag{4}$$

where $k_{F_i}$ is the Fermi momentum of species $i$. The two-body cluster energy $E_2$ accounts for correlations and interactions and can be written as:

$$E_2 = \frac{1}{2A}\sum_{i<j}\langle ij|W(12)|\ ij - ji\rangle, \tag{5}$$

where

$$W(12) = [f(12), [T_{12}, f(12)]] + f(12)v(12)f(12) \tag{6}$$

with $f(12)$ the two-body correlation function, $T_{12}$ is two-body kinetic energy and the particle state i.e., $|ij\rangle$ is defined as:

$$|ij\rangle = |k_i k_j,\ \sigma_1\sigma_j,\ \tau_i\tau_j\rangle, \tag{7}$$

for asymmetric nuclear matter and in case of including hyperons (LOCVY), the particle state should include $s_i$, $s_j$ too:

$$|ij\rangle = |k_i k_j,\ \sigma_1\sigma_j,\ \tau_i\tau_j,\ s_i s_j\rangle. \tag{8}$$

The above quantum numbers stand for the wavenumber, the spin state, the isospin state, and the strangeness number of two particles, respectively.

The correlation function $f(12)$ is expanded over operator channels labeled by quantum numbers $\{J, L, S, T, T_z\}$ and includes tensor terms via operator basis functions:

$$f(ij) = \sum_{\alpha,p=1}^{3} f_\alpha^p(ij) O_\alpha^p(ij), \tag{9}$$

where $O_\alpha^p$ are the projection operators onto the channels labeled by $\alpha = \{J, L, S, T, T_z\}$, with $p = 1$ corresponding to uncoupled channels (singlet or triplet states with $J = L$ and $p = 2$, 3 corresponding to coupled channels with $(J = L \pm 1)$). The operators $O_\alpha^p$ are expressed as follows:

$$O_\alpha^{=1-3} = 1, \left(\frac{2}{3} + \frac{1}{6}S_{12}\right), \left(\frac{1}{3} - \frac{1}{6}S_{12}\right). \tag{10}$$

where $S_{12} = 3(\sigma_1.\hat{r})(\sigma_2.\hat{r}) - \sigma_1.\sigma_2$ is the tensor operator. The correlation functions are obtained by solving the associated Euler–Lagrange equations, subject to a normalization constraint.

To maintain computational efficiency, the cluster expansion is truncated at the two-body level. However, the energy of three-body clusters is effectively suppressed by choosing correlation functions that minimize the functional energy under the normalization constraint; the contribution of higher clusters is rendered negligible (H. Moshfegh & M. Modarres 1998). Nevertheless, TBF can still be included explicitly in the Hamiltonian.

The energy expression is now minimized with respect to the channel-dependent correlation functions, subject to the following normalization constraint (M. Modarres & H. Moshfegh 2000; M. Modarres & A. Tafrihi 2023):

$$\frac{1}{A}\sum_{ij}\langle ij|f_P^2(12) - f^2(12)|ij - ji\rangle = 0. \tag{11}$$

In addition, the correlation functions are required to "heal" to the Pauli function $f_P(r)$ at large distances:

$$f_P(r) = \begin{cases} \left[1 - \frac{9}{2}\left(\frac{j_1(k_{F_i}r)}{k_{F_i}r}\right)^2\right]^{-1/2}, & \text{indistinguishable pair} \\ 1, & \text{distinguishable pair} \end{cases}, \tag{12}$$

where $j_1(x)$ is the spherical Bessel function of the first kind of order one. The normalization constraint introduces Lagrange multiplier parameters into the formalism.

Minimizing Equation (5) yields a set of Euler–Lagrange differential equations for the correlation functions $f_\alpha(ij)$ in each spin–isospin–angular momentum channel $\alpha$. Solving these equations determines the correlation functions and hence the two-body cluster energy.

### 2.2. *Interactions*

Accurate modeling of baryon–baryon interactions is fundamental to any microscopic calculation of dense matter. In the LOCVY formalism, both the correlation functions and the energy expectation value depend explicitly on the underlying interaction potentials. The total hypernuclear Hamiltonian is decomposed into nucleon–nucleon (NN), hyperon–nucleon (YN), and hyperon–hyperon (YY) sectors, each contributing to the structure and energetics of the system.

The choice of two-body and three-body interaction models directly influences the predicted EOS and composition of dense baryonic matter. Therefore, employing phenomenology-based, data-constrained potentials is crucial to ensure consistency with empirical nuclear and hypernuclear data, including binding energies, scattering phase shifts, and hypernuclear structure observables.

#### 2.2.1. *Nucleon–Nucleon Interaction*

For the NN sector, including $V_{nn}$, $V_{pp}$, and $V_{np}$ interactions, we employ the high-precision Argonne $V_{18}$ (AV18) potential (R. B. Wiringa et al. 1995). This two-body potential is widely used in nuclear structure and nuclear matter calculations due to its excellent reproduction of NN scattering data and deuteron properties.

The AV18 potential is constructed as an operator expansion:

$$v(ij) = \sum_{\alpha=1}^{18} v_\alpha(ij)\, O^\alpha(ij), \tag{13}$$

where the first 14 operators are charge-independent and include central, spin, isospin, tensor, and spin–orbit terms:

$$O^{\alpha=1-14}(ij) = 1,\ \boldsymbol{\tau}_i \cdot \boldsymbol{\tau}_j,\ \boldsymbol{\sigma}_i \cdot \boldsymbol{\sigma}_j,\ (\boldsymbol{\sigma}_i \cdot \boldsymbol{\sigma}_j)(\boldsymbol{\tau}_i \cdot \boldsymbol{\tau}_j),\ S_{ij},\ S_{ij}(\boldsymbol{\tau}_i \cdot \boldsymbol{\tau}_j),\ \boldsymbol{L} \cdot \boldsymbol{S},\ (\boldsymbol{L} \cdot \boldsymbol{S})(\boldsymbol{\tau}_i \cdot \boldsymbol{\tau}_j),\ L^2,\ L^2(\boldsymbol{\tau}_i \cdot \boldsymbol{\tau}_j),\ L^2(\boldsymbol{\sigma}_i \cdot \boldsymbol{\sigma}_j),\ L^2(\boldsymbol{\sigma}_i \cdot \boldsymbol{\sigma}_j)(\boldsymbol{\tau}_i \cdot \boldsymbol{\tau}_j),\ (\boldsymbol{L} \cdot \boldsymbol{S})^2,\ (\boldsymbol{L} \cdot \boldsymbol{S})^2 (\boldsymbol{\tau}_i \cdot \boldsymbol{\tau}_j). \quad (14)$$

The remaining four operators explicitly break charge independence:

$$O^{\alpha=15-18}(ij) = T_{ij},\ (\boldsymbol{\sigma}_i \cdot \boldsymbol{\sigma}_j)T_{ij},\ S_{ij}T_{ij},\ (\tau_{zi} + \tau_{zj}), \quad (15)$$

where the isotensor operator $T_{ij} = 3\tau_{zi}\tau_{zj} - \boldsymbol{\tau}_i \cdot \boldsymbol{\tau}_j$ is defined analogously to the tensor operator $S_{ij}$.

Although the AV18 potential accurately captures two-body correlations, calculations based solely on two-body forces cannot reproduce the empirical saturation point of symmetric nuclear matter. To address this deficiency, we include a TBF using the semiphenomenological Urbana IX model (Z. Li & H.-J. Schulze 2008). Rather than solving the full three-body problem explicitly, the TBF is incorporated as an effective, density-dependent two-body interaction. This is achieved by averaging over the third nucleon, weighted by the LOCV two-body correlation functions at a given baryon density (S. Goudarzi et al. 2018; S. Goudarzi & H. R. Moshfegh 2015).

### 2.2.2. ΛN Interaction

For the $\Lambda N$ interaction, several meson-exchange models have been proposed based on SU(3) flavor symmetry and baryon–meson coupling constants. These theoretical models, such as the Nijmegen soft-core potentials, are all compatible with implementation in the LOCV framework(M. Shahrbaf et al. 2019). In this work, we adopt a spin-parity-dependent potential, whose central part is expressed as a three-range Gaussian form, as given in Equation (16) (E. Hiyama et al. 2006, 2009). This potential effectively reproduces the high-energy $\Lambda N$ phase shifts obtained from the NSC97f model (T. A. Rijken et al. 1999).

To further improve agreement with experimental hypernuclear data, specific components of the potential are tuned. The strength of the second Gaussian range in the even-parity triplet and singlet channels is adjusted to reproduce the energy splitting between the spin-doublet states $1^+$ and $0^+$ in ${}^4_\Lambda$H (H. Togashi et al. 2016). Similarly, the odd-parity sector is constrained by fitting the separation energy of ${}^4_\Lambda$Li (H. Togashi et al. 2016). The resulting potential is written as:

$$V_{\Lambda N}^{(c)}(r) = \sum_{\alpha}\sum_{i=1}^{3} v_i^{\alpha} \exp\left[-\left(\frac{r}{\beta_i}\right)^2\right], \quad (16)$$

where $\alpha$ labels the four spin-parity channels: singlet-even (${}^1E$), triplet-even (${}^3E$), singlet-odd (${}^1O$), and triplet-odd (${}^3O$), and $r$ is the relative coordinate between the $\Lambda$ and the nucleon.

The range parameters $\beta_i$ and the strengths $v_i^{\alpha}$ for each spin-parity channel are given in Table 1.

### 2.2.3. ΛΛ Interaction

To account for the presence of multiple $\Lambda$ hyperons in dense baryonic matter, it is essential to include a data-constrained $\Lambda\Lambda$ interaction in the hypernuclear Hamiltonian. Analogous to the $\Lambda N$ case, we employ a central, spin-dependent, three-range Gaussian potential for both the even-parity and odd-parity components of the $\Lambda\Lambda$ interaction.

**Table 1**
Parameters of the $\Lambda N$ Interaction Used in Equation (16) (M. Shahrbaf et al. 2019)

| $i$ | 1 | 2 | 3 |
|---|---|---|---|
| $\beta_i$ (fm) | 1.60 | 0.80 | 0.35 |
| $v_i({}^1\mathrm{E})$ (MeV) | −7.87 | −357.4 | 6132.0 |
| $v_i({}^3\mathrm{E})$ (MeV) | −7.89 | −217.3 | 3139.0 |
| $v_i({}^1\mathrm{O})$ (MeV) | −1.30 | 513.7 | 8119.0 |
| $v_i({}^3\mathrm{O})$ (MeV) | −3.38 | 22.9 | 5952.0 |

**Table 2**
Parameters for the Even-parity Component of the $\Lambda\Lambda$ Potential in Equation (17) (M. Shahrbaf et al. 2019)

| $i$ | 1 | 2 | 3 |
|---|---|---|---|
| $\mu_i^{\mathrm{even}}$ (fm$^{-2}$) | 0.555 | 1.656 | 8.163 |
| $\nu_i^{\mathrm{even}}$ (MeV) | −10.67 | −93.51 | 4884.0 |
| $\nu_i^{\sigma,\mathrm{even}}$ (MeV) | 0.0966 | 16.08 | 915.8 |

The parameters of the even-state potential are constrained by available experimental data from double-$\Lambda$ hypernuclei, most notably the observed binding energy of the $S$-wave state in ${}^{6}_{\Lambda\Lambda}$He (E. Hiyama et al. 2002). Since all confirmed double-$\Lambda$ states involve two hyperons in relative $S$-orbitals (i.e., $L = 0$), the even-parity sector is more reliably constrained. For this reason, the even-state potential is calibrated to reproduce the low-energy features of the $\Lambda\Lambda$ channel in the Nijmegen model F potential (T. A. Rijken et al. 2010), using Gaussian terms of the form:

$$V_{12}^{\Lambda\Lambda,\,\mathrm{even}}(r) = \sum_{i=1}^{3}(\nu_i^{\mathrm{even}} + \nu_i^{\sigma,\mathrm{even}}\ \sigma_1 \cdot \sigma_2) \times \exp\left[-\mu_i^{\mathrm{even}} r^2\right], \quad (17)$$

where $\nu_i^{\mathrm{even}}$ and $\nu_i^{\sigma,\mathrm{even}}$ denote the central and spin-dependent strengths of the Gaussian components, respectively, and $\mu_i^{\mathrm{even}}$ determines their range. The explicit parameters are listed in Table 2.

Due to the scarcity of experimental constraints on the odd-parity $\Lambda\Lambda$ interaction, arising from the absence of observed $P$-wave $\Lambda\Lambda$ bound states, its modeling remains somewhat uncertain. Nevertheless, we assume the same functional form as the even-state interaction and adopt parameters adjusted to maintain consistency with hypernuclear structure models:

$$V_{12}^{\Lambda\Lambda,\,\mathrm{odd}}(r) = \sum_{i=1}^{3}(\nu_i^{\mathrm{odd}} + \nu_i^{\sigma,\mathrm{odd}}\ \sigma_1 \cdot \sigma_2) \times \exp\left[-\mu_i^{\mathrm{odd}} r^2\right]. \quad (18)$$

The odd-parity parameters, which are chosen to explore the attractive limit of $\Lambda\Lambda$ interactions, are the same as those presented in Table 2. This phenomenological modeling of the $\Lambda\Lambda$ interaction, while constrained primarily by limited experimental data, provides a workable input for our LOCV-based EOS calculations. It allows us to assess the role of $\Lambda\Lambda$ correlations (M. Shahrbaf et al. 2019) in hyperonic matter and their impact on the structure of NS.

## 3. Equation of State for Hypernuclear Matter

Understanding the EOS of dense baryonic matter is crucial for determining the structure and composition of NSs, especially in their inner cores, where extreme densities may allow for the appearance of strange particles such as hyperons. The EOS not only governs the mass–radius relation of compact stars but also sets constraints on their maximum mass, internal composition, and cooling behavior. In this section, we formulate the EOS of hypernuclear matter under conditions relevant to NSs.

### 3.1. β-stable Matter

NS matter is governed by weak interaction processes that drive the system toward $\beta$ equilibrium. This equilibrium is established under three primary constraints: charge neutrality, baryon number conservation, and chemical equilibrium among all particle species involved. Leptons (electrons and muons) are included to balance charge neutrality, while heavier leptons, such as $\tau$, are neglected due to their large rest mass.

At zero temperature, the neutrino mean free path exceeds the stellar radius, allowing one to neglect trapped neutrinos (S. L. Shapiro & S. A. Teukolsky 1983). The weak processes that establish $\beta$ equilibrium include:

$$n \rightarrow p + e + \bar{\nu}_e, \tag{19}$$

$$p + e \rightarrow n + \nu_e. \tag{20}$$

These lead to the equilibrium condition:

$$\mu_n = \mu_p + \mu_e. \tag{21}$$

When the electron chemical potential exceeds the muon rest mass, muon production becomes energetically favorable:

$$\mu_e = \mu_\mu. \tag{22}$$

At even higher densities, strange baryons such as the neutral Lambda ($\Lambda$) hyperon may appear through reactions like:

$$n + n \rightarrow n + \Lambda, \tag{23}$$

which implies the condition:

$$\mu_n = \mu_\Lambda. \tag{24}$$

The charge neutrality and baryon number conservation conditions read:

$$\rho_p = \rho_e + \rho_\mu, \tag{25}$$

$$\rho_b = \rho_n + \rho_p + \rho_\Lambda. \tag{26}$$

Here, $\mu_i$ and $\rho_i$ denote the chemical potential and number density of particle species $i$, respectively.

The total energy density of baryonic matter is given by:

$$\epsilon_b = \rho_b\left(\frac{E}{A}\right), \tag{27}$$

where $E/A$ is the energy per particle, obtained from the LOCVY method including kinetic and interaction terms. The baryonic chemical potentials are computed via partial derivatives of the energy density:

$$\mu_n(\rho_b, x_p, x_\lambda) = \frac{\partial \epsilon_b}{\partial \rho_n}, \tag{28}$$

$$\mu_p(\rho_b, x_p, x_\lambda) = \frac{\partial \epsilon_b}{\partial \rho_p}, \tag{29}$$

$$\mu_\lambda(\rho_b, x_p, x_\lambda) = \frac{\partial \epsilon_b}{\partial \rho_\lambda}, \tag{30}$$

where $x_i = \frac{\rho_i}{\rho_b}$, $(\sum_{i=p,n,\Lambda} x_i = 1)$.

Since leptons are treated as relativistic Fermi gases, their chemical potentials correspond to their total relativistic Fermi energies at zero temperature. The analytic form of the chemical potential $\mu_l$ for a lepton $l = e, \mu$ is given by:

$$\mu_l = (\hbar^2 c^2 (3\pi^2 \rho_l)^{\frac{2}{3}} + m_l^2 c^4)^{\frac{1}{2}}. \tag{31}$$

Depending on the density regime, the composition of matter evolves as follows:

(i) *Low-density regime (no muons, no hyperons):–*

$$\mu_n = \mu_p + \mu_e, \tag{32}$$

$$\rho_p = \rho_e, \tag{33}$$

$$\rho_b = \rho_n + \rho_p. \tag{34}$$

(ii) *Intermediate-density regime (muons present):*

$$\mu_n = \mu_p + \mu_e, \tag{35}$$

$$\mu_e = \mu_\mu, \tag{36}$$

$$\rho_p = \rho_e + \rho_\mu, \tag{37}$$

$$\rho_b = \rho_n + \rho_p. \tag{38}$$

(iii) *High-density regime (muons and $\Lambda$ present):*

$$\mu_n = \mu_p + \mu_e, \tag{39}$$

$$\mu_e = \mu_\mu, \tag{40}$$

$$\mu_n = \mu_\Lambda, \tag{41}$$

$$\rho_p = \rho_e + \rho_\mu, \tag{42}$$

$$\rho_b = \rho_n + \rho_p + \rho_\Lambda. \tag{43}$$

These coupled equations are solved self-consistently at each total baryon density $\rho_b$ to determine the composition of $\beta$-stable hyperonic matter. The appearance of each new particle species defines a critical threshold density, beyond which it contributes to the EOS and composition of the system.

### 3.2. Particle Fractions and Chemical Potentials

Understanding the evolution of particle fractions and chemical potentials in dense matter is essential for modeling the physics of neutron and hyperonic stars. Once the coupled equations governing $\beta$-equilibrium, charge neutrality, and baryon number conservation are solved, the equilibrium composition of the system is fully determined. We define the particle fractions as

$$x_i = \frac{\rho_i}{\rho_b}, \quad i = n, p, e, \mu, \Lambda.$$

Figure 1 displays the chemical potentials of neutrons, protons, lambda, and electrons as functions of the baryon density under conditions of $\beta$-equilibrium. Dashed lines represent the pure nucleonic matter scenario, while solid lines correspond to the composition including $\Lambda$ hyperons. The green dotted line represents the chemical potential of the $\Lambda$ particles. The threshold density at which the $\Lambda$ particle first appears is marked by a small yellow circle, beyond which the $\beta$-equilibrium condition implies $\mu_\Lambda = \mu_n$. The figure reveals three distinct density regimes, marking the sequential appearance of leptons (muons) and hyperons ($\Lambda$). Notably, due to the $\beta$-equilibrium conditions, the chemical potential of muons becomes equal to that of electrons after muon onset (at around $\rho_b = 0.17\ \mathrm{fm}^{-3}$ in our model), and similarly, the chemical potentials of neutrons and $\Lambda$ hyperons converge following hyperon appearance (at around $\rho_b = 0.38\ \mathrm{fm}^{-3}$ in our model). As evident from the figure, the inclusion of the $\Lambda$ hyperon leads to a reduction in the chemical potentials of all particle species at a given baryon density. This decrease reflects the redistribution of baryon number among additional degrees of freedom introduced by the hyperons, thereby lowering the overall energy cost of particle addition and softening the EOS. This intricate interplay between species profoundly impacts the macroscopic properties of NSs, such as their mass–radius relationship and stability.

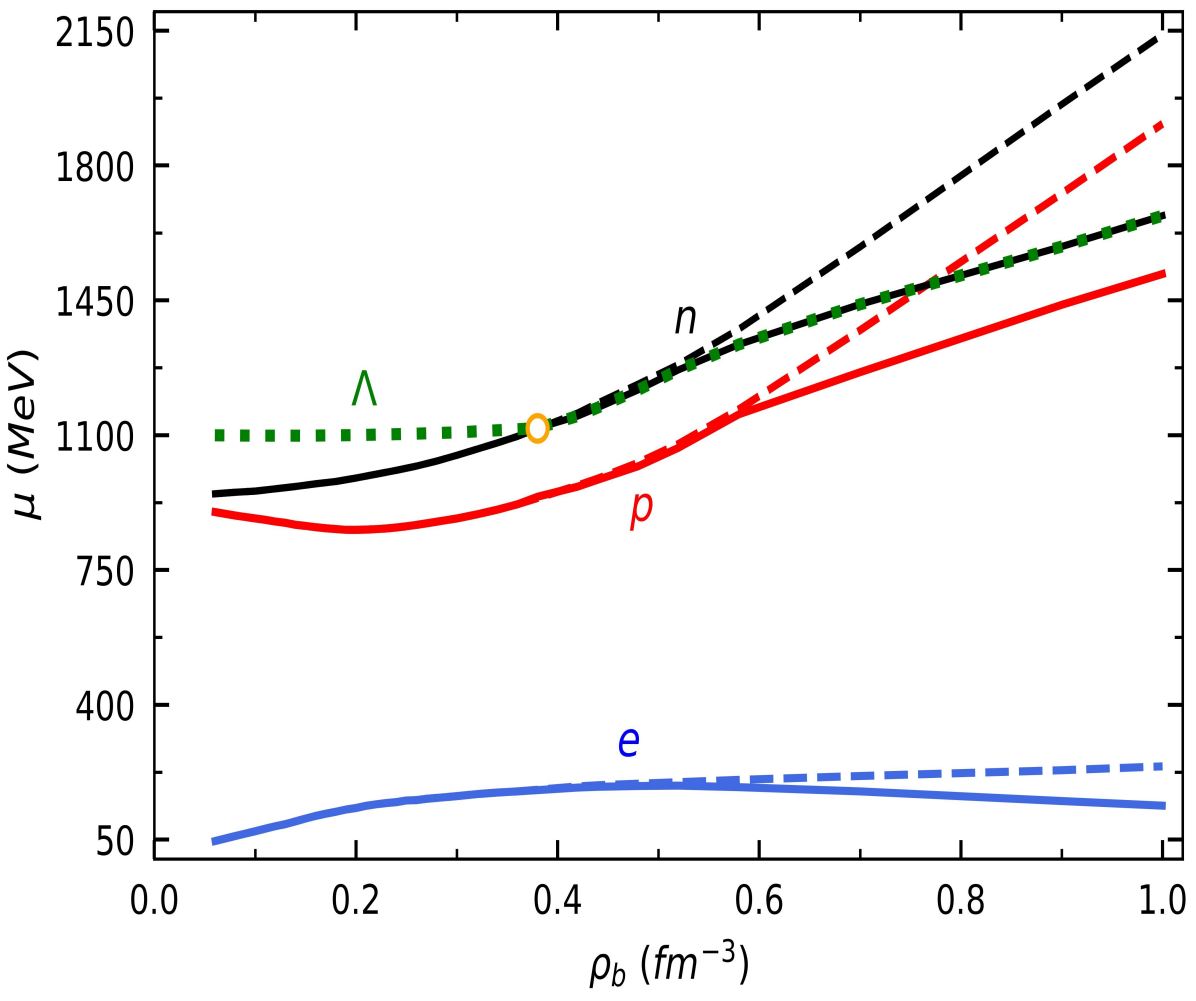

**Figure 1.** Chemical potentials of neutrons, lambda, protons, and electrons as functions of the baryon density under $\beta$-equilibrium conditions. Dashed lines represent the case of pure nucleonic matter, while solid lines correspond to matter including $\Lambda$ hyperons. The green dotted line shows the $\Lambda$-hyperon chemical potential, and the $\Lambda$ onset density is indicated by a small yellow open circle.

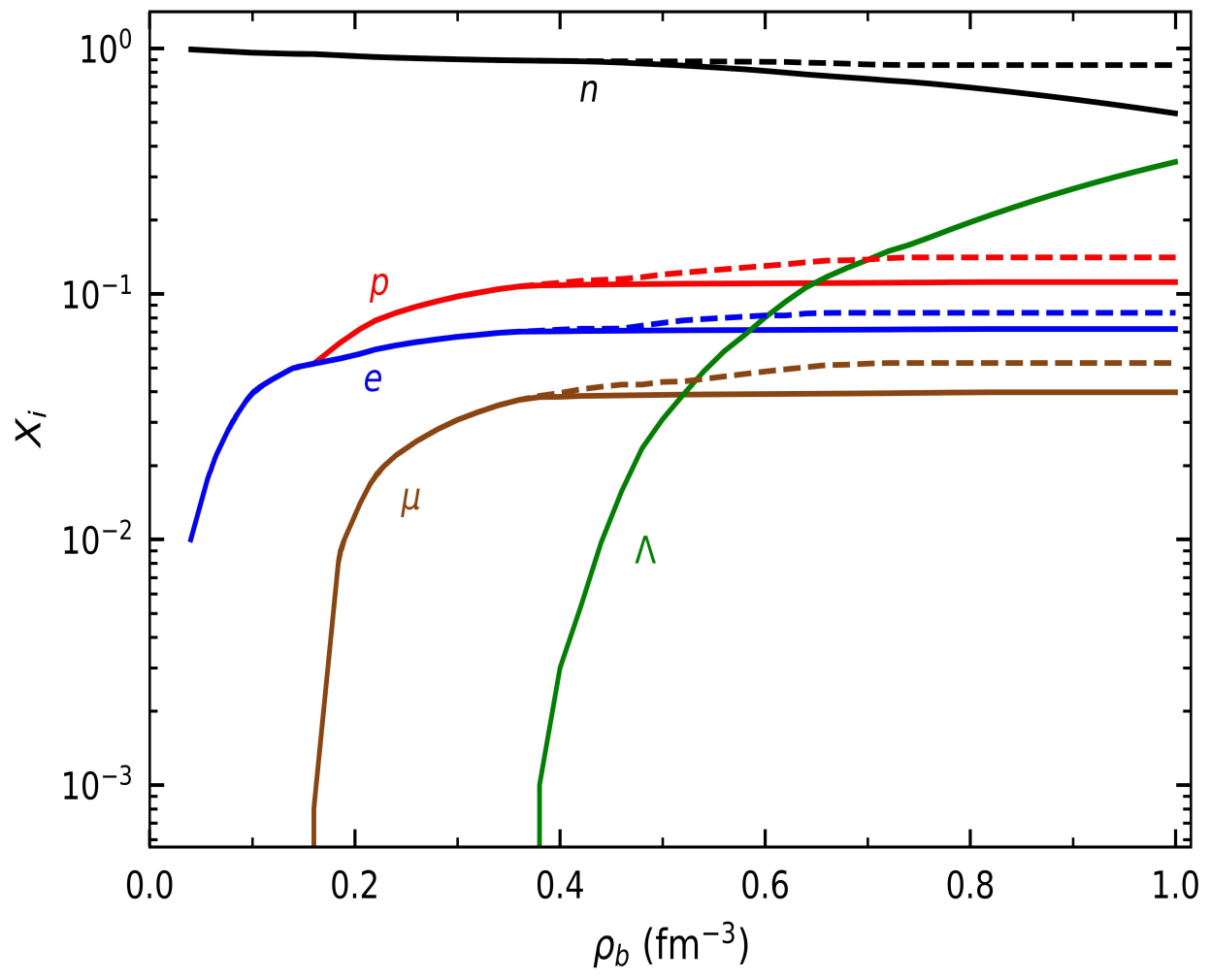

**Figure 2.** Fractional abundances of particles as functions of the baryon density under conditions of $\beta$-equilibrium. Dashed curves correspond to the pure nucleonic case, while solid curves represent the composition after the onset of the $\Lambda$ hyperon.

Figure 2 displays the particle fractions as a function of the baryon density. At low densities, the matter is mainly composed of neutrons and a small fraction of protons and electrons. Once the electron chemical potential exceeds the muon rest mass, muons appear at a threshold density around $\rho_b \approx 0.17\ \mathrm{fm}^{-3}$. As the density increases further, the neutron chemical potential eventually becomes sufficient to produce $\Lambda$ hyperons, which appear at a threshold of $\rho_b \approx 0.38\ \mathrm{fm}^{-3}$. Beyond this point, the dashed lines, corresponding to matter without hyperons, are no longer valid, and only the solid lines, which represent the composition including $\Lambda$ hyperons, remain applicable. The emergence of the $\Lambda$ hyperon causes a significant decrease in all particle fractions, particularly the neutron and proton fractions. Therefore, the system readjusts to maintain charge neutrality and $\beta$-equilibrium, with $\Lambda$ assuming an increasing share of the baryon number at higher densities. This redistribution softens the EOS by lowering the pressure at high densities, a feature consistent with predictions from BHF and RMF approaches (I. Vidana et al. 2011; S. Weissenborn et al. 2012).

Figure 3 compares the $\Lambda$-hyperon fraction obtained within the LOCVY framework with those predicted by various other ab initio approaches and RMF models. It illustrates how the onset and the abundance of $\Lambda$ particles vary with increasing baryon density. One can see in the figure that, in our approach, which does not include a TBF for the $\Lambda$ hyperon, the onset density of $\Lambda$ is higher compared to some other approaches that primarily incorporate hyperonic TBF. Additionally, the $\Lambda$ fraction in our model increases more gradually after its emergence. This behavior is the main reason why, even with hyperons included, the hyperon puzzle does not manifest strongly, and the maximum mass of the star can still reach $2\,M_\odot$. Indeed, at high densities, the repulsive character of the $\Lambda\Lambda$ interaction reduces the softening associated with hyperon onset and helps the EOS support $\approx 2\,M_\odot$. Furthermore, in the LOCV framework, explicit two-body, channel-dependent correlation functions provide additional medium-induced repulsion, while the nucleonic TBF stiffens the nucleonic sector and delays hyperon appearance. These combined effects moderate the overall softening once hyperons are present. However, a more precise determination of Λ binding energies in single- and double-Λ hypernuclei will require the explicit inclusion of hyperonic TBF of the types $\Lambda NN$ and $\Lambda\Lambda N$, as well as a possible retuning of the ΛΛ interactions. The corresponding baryon density thresholds for the appearance of Λ hyperons in each model are listed in Table 3, providing a quantitative measure of the conditions under which hyperons first emerge in dense matter.

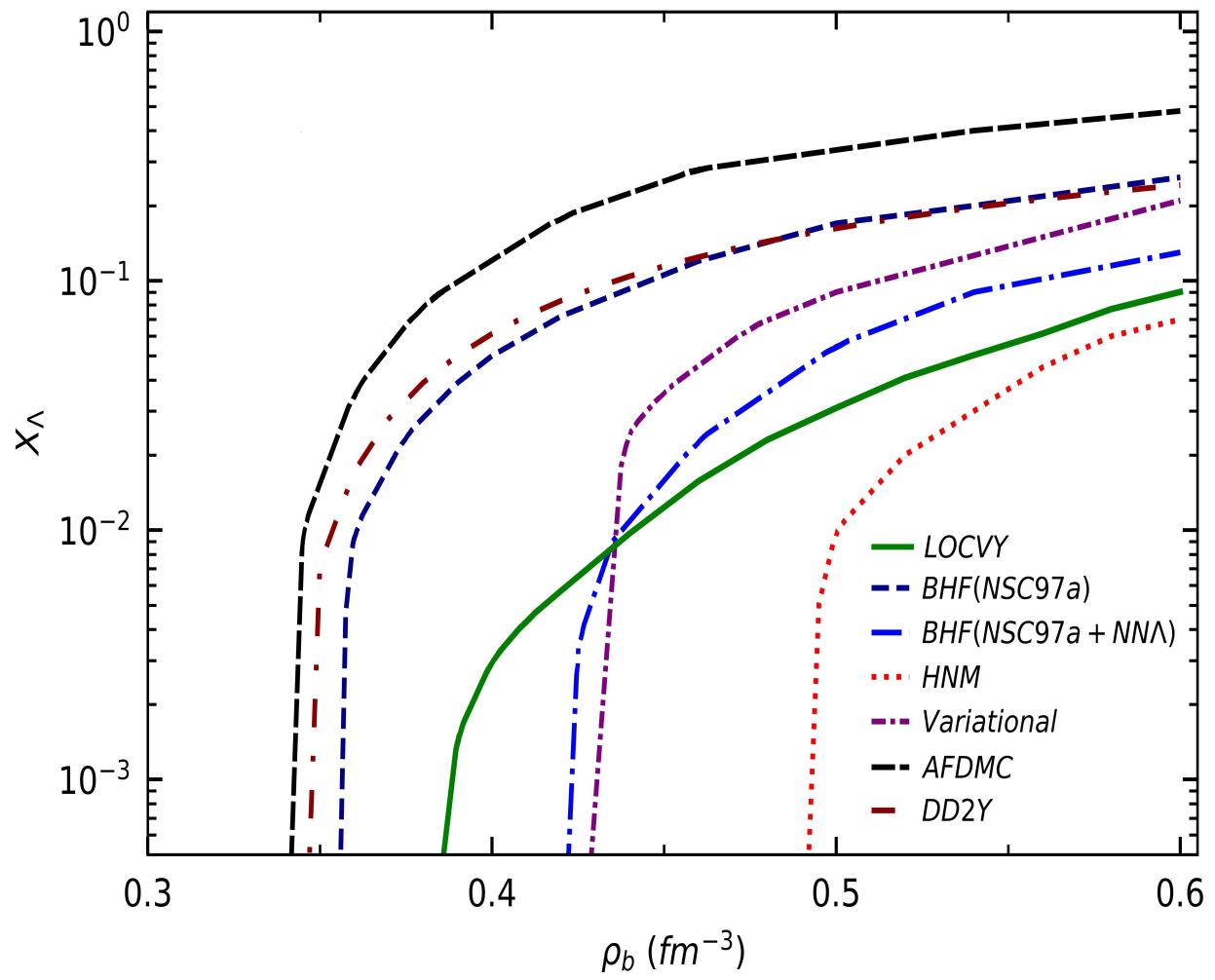

**Figure 3.** $\Lambda$ fraction as a function of the baryon density within the LOCVY framework compared with results obtained from various other approaches, including: lattice EFT using two-body and three-body hyperonic forces HNM (H. Tong et al. 2025c); the BHF approach with and without hyperonic TBFs (H. Tong et al. 2025a); AFDMC with hyperonic TBFs (D. Lonardoni et al. 2015); the variational method with hyperonic TBFs (H. Togashi et al. 2016); and finally, an RMF approach including the $\Phi$ meson and only the $\Lambda$ hyperon (DD2Y-T; J. R. Stone et al. 2021).

**Table 3**
Threshold Baryon Density for Appearance of the Λ Hyperon in Different Models

| Method | Threshold Density of Λ ($fm^{-3}$) |
|---|---|
| LOCVY (present work) | 0.38 |
| AFDMC (D. Lonardoni et al. 2015) | 0.33 |
| BHF (NSC97a) (H. Tong et al. 2025a) | 0.35 |
| BHF (NSC97a+NNΛ) (H. Tong et al. 2025a) | 0.41 |
| HNM (H. Tong et al. 2025c) | 0.49 |
| Variational (H. Togashi et al. 2016) | 0.42 |
| DD2Y-T (J. R. Stone et al. 2021) | 0.34 |

### 3.3. Equation of State of β-stable Hypernuclear Matter

Once the particle fractions and chemical potentials at different baryon densities are determined, the next crucial step is to construct the EOS for $\beta$-equilibrated matter. The EOS plays a pivotal role in describing the internal structure and global properties of compact objects such as NSs and hyperonic stars.

The total energy density of the system is composed of two primary contributions: the baryonic energy density $\epsilon_b$ (including nucleons and Λ hyperons) and the leptonic energy density $\epsilon_l$:

$$\varepsilon = \epsilon_b + \epsilon_l. \tag{44}$$

In our model, leptons are treated as a relativistic, noninteracting Fermi gas. At zero temperature, the leptonic contribution is given by

$$\epsilon_l = \sum_{l=e,\mu} \frac{2}{h^3} \int_0^{P_{F_l}} d^3p_l \sqrt{(p_l c)^2 + (m_l c^2)^2}, \tag{45}$$

which can be analytically expressed as

$$\epsilon_l = \sum_{l=e,\mu} \frac{\pi c^5 m_l^4}{h^3} [\phi_l (2\phi_l^2 + 1)\sqrt{\phi_l^2 + 1} - \sinh^{-1}\phi_l], \tag{46}$$

where $\phi_l = \frac{P_{F_l}}{m_l c}$ and $P_{F_l}$ is the Fermi momentum of lepton $l$.

The baryonic pressure is computed from the derivative of the energy per particle:

$$P_b = \rho_b^2 \frac{\partial}{\partial \rho_b}\left(\frac{E}{A}\right), \tag{47}$$

while the pressure of the relativistic leptons is obtained thermodynamically as:

$$P_l = -\epsilon_l + \sum_{l=e,\mu} \rho_l \mu_l. \tag{48}$$

The total pressure is then:

$$P = P_l + P_b. \tag{49}$$

As illustrated in Figure 4, the inclusion of Λ hyperons induces a substantial modification in the thermodynamic behavior of dense matter. At high baryon densities, the presence of hyperons provides an additional degree of freedom, allowing the system to redistribute its energy among more particle species. Consequently, the energy per baryon (the upper panel of Figure 4) decreases relative to the purely nucleonic case. A more striking effect is observed in the pressure profile (the lower panel of Figure 4), which becomes noticeably softer once hyperons appear.

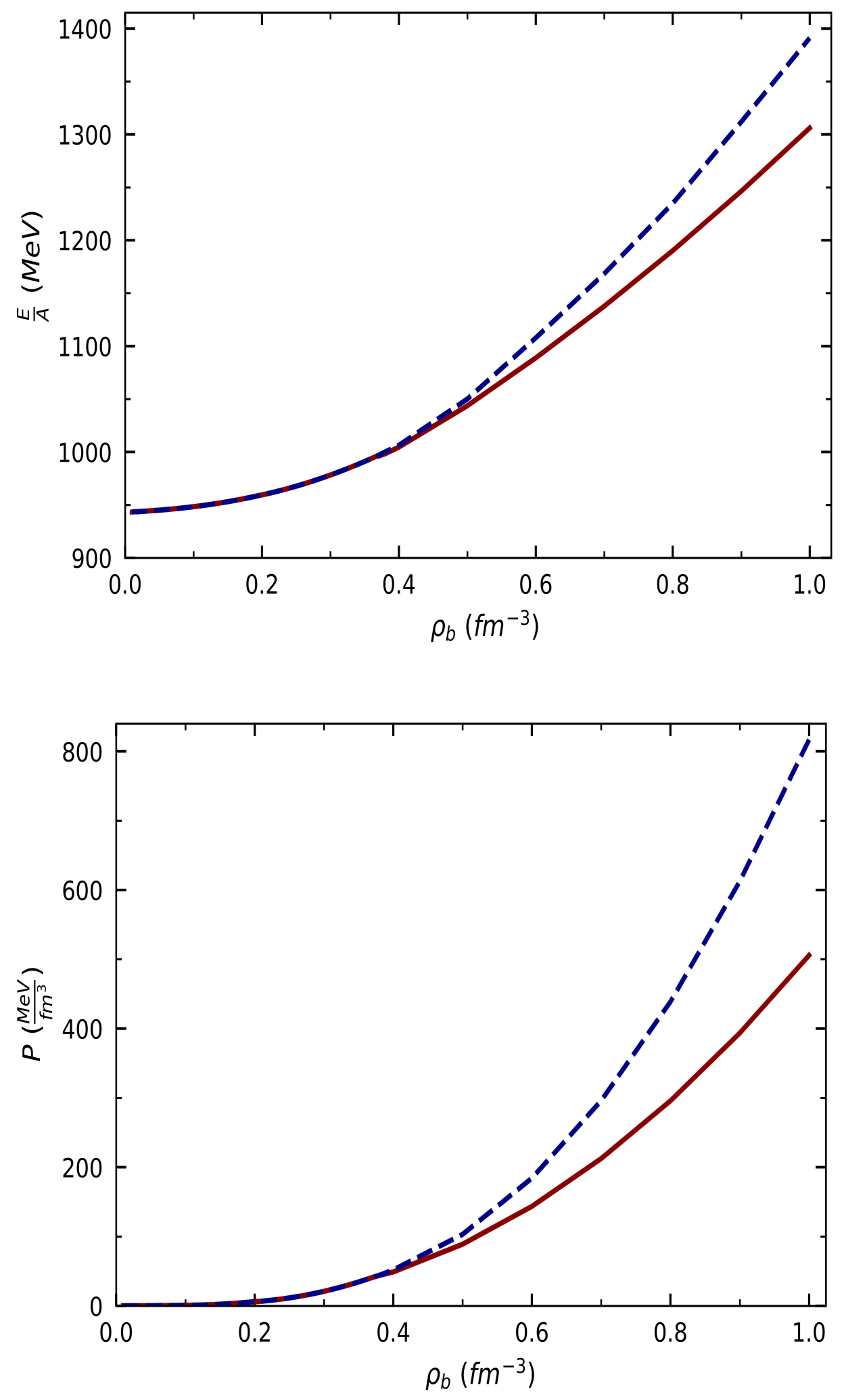

**Figure 4.** Top panel: energy per baryon as a function of the baryon density, with (solid curve) and without (dashed curve) Λ hyperons. Bottom panel: corresponding pressure for the two cases.

This softening, a well-established consequence of hyperon onset in dense matter (J. Schaffner-Bielich 2010; I. Bombaci 2017), arises from the fact that hyperons are more massive than nucleons and begin to populate the system at high densities. Since they can occupy lower momentum states and do not contribute as strongly to the Fermi pressure, they partially replace high-momentum nucleons, leading to a reduction in the overall pressure at a given energy density. Such a reduction in stiffness of the EOS has direct astrophysical consequences. In particular, it influences the mass–radius relation and leads to a lower maximum mass for NSs, which is a central aspect of the so-called hyperon puzzle, a point that will be examined in detail in the following section.

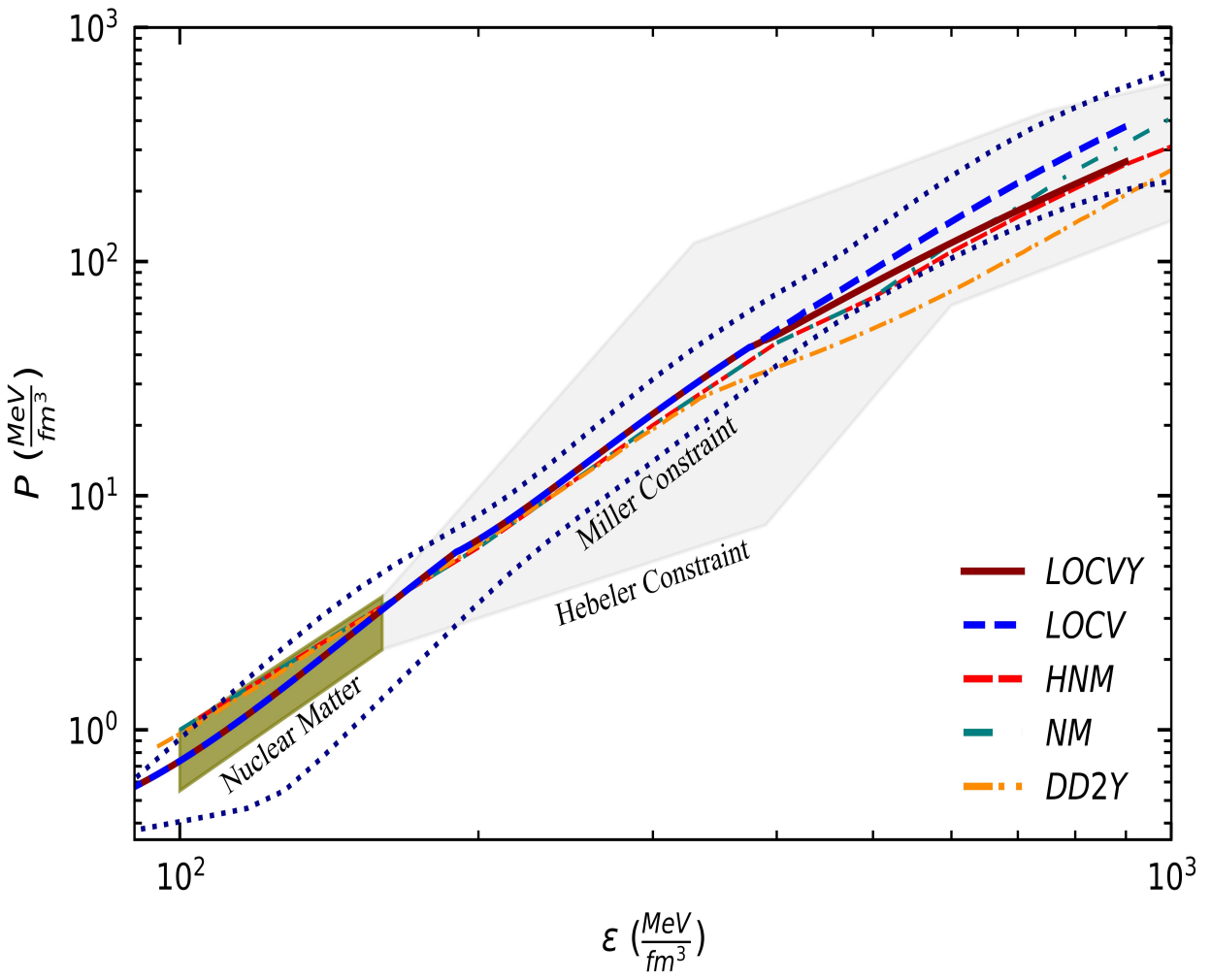

**Figure 5.** Pressure as a function of the energy density for the obtained EOSs, with and without hyperons. Our nucleonic EOS is represented by a blue dashed line, while the hyperonic EOS is shown as a red solid line. For comparison, the figure also includes the nucleonic EOS (NM) and hyperonic EOS (HNM) from H. Tong et al. (2025c), as well as the DD2Y-T EOS from J. R. Stone et al. (2021), which incorporates only the Λ hyperon. The gray-shaded region denotes the EOS constraint from K. Hebeler et al. (2013), while the narrow band between blue dotted lines corresponds to the constraint from the analysis in M. C. Miller et al. (2019).

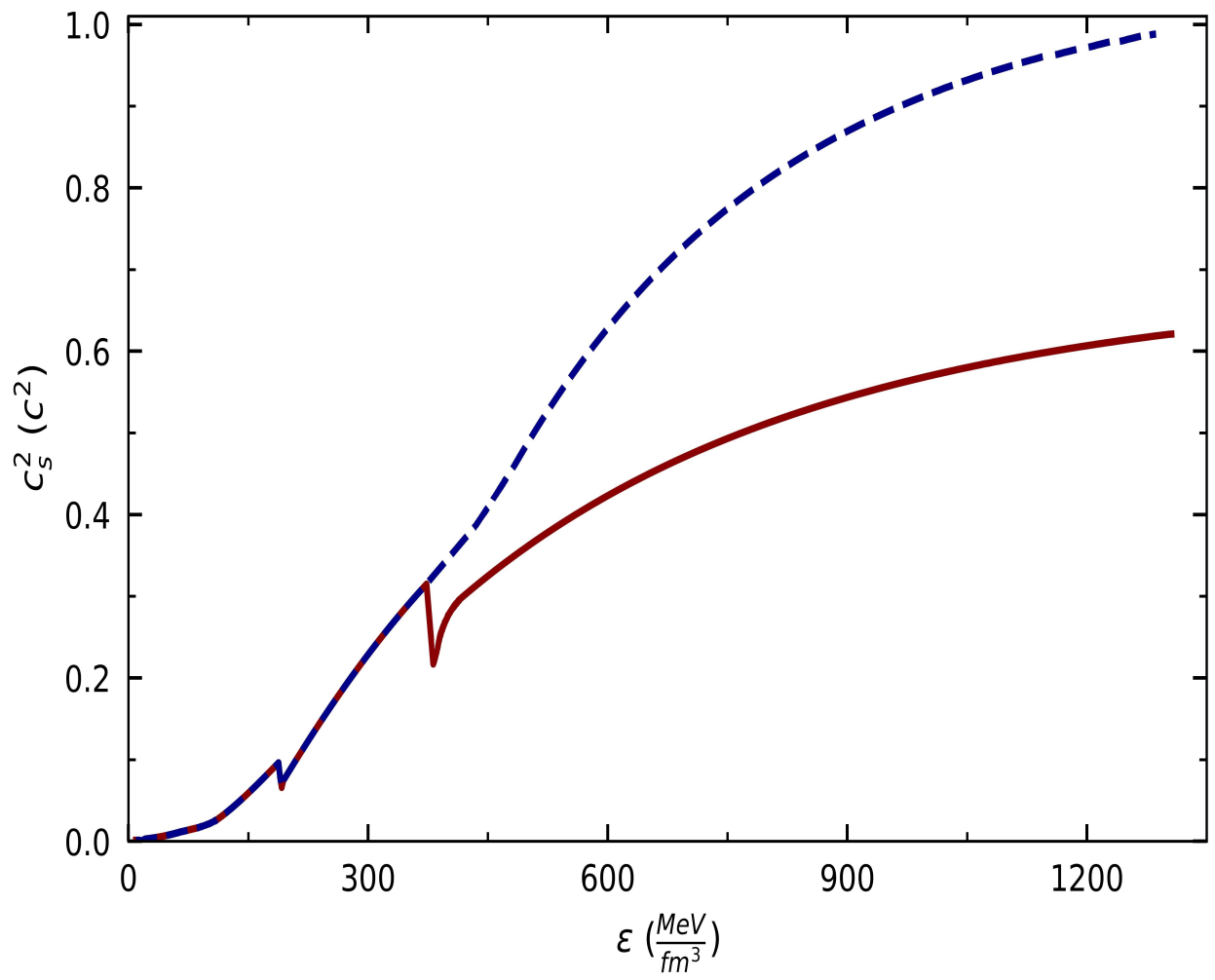

**Figure 6.** Squared speed of sound for the obtained EOSs: nucleonic matter (dashed curve) and hyperonic matter (solid curve).

Figure 5 presents the EOS results in the pressure–energy density plane for both the LOCV (nuclear matter only) and LOCVY (including Λ hyperon) models. In this figure, our results are compared with other nucleonic and hyperonic EOSs taken from H. Tong et al. (2025c) and J. R. Stone et al. (2021). The HNM EOSs are derived using lattice EFT, while DD2Y-T is based on an RMF approach. All of the hyperonic EOSs considered include only the Λ hyperon. In this figure, the obtained $P$–$\varepsilon$ relations for both of our EOSs not only lie entirely within the gray band, which represents the accepted range of pressure–energy density curves reported by Hebeler et al. in K. Hebeler et al. (2013), but also pass well through the narrower region bounded by blue dotted lines, which corresponds to the EOS constraint from M. C. Miller et al. (2019). This agreement demonstrates that our microscopic LOCV framework, both with and without hyperons, remains compatible with empirical and theoretical constraints derived from chiral EFT and astrophysical observations.

Figure 6 shows the squared speed of sound, $c_s^2 = \frac{dP}{d\varepsilon}$, for the nucleonic and hyperonic EOSs obtained in this work. This quantity characterizes the stiffness of the EOS and the response of dense matter to compression. In NS matter, it provides crucial insight into the underlying microphysics, such as the appearance of new degrees of freedom. Deviations from the conformal limit, $c_s^2 = 1/3$, which is expected at very high densities, may signal nonperturbative effects or the onset of exotic phases such as quark deconfinement. As illustrated in the figure, at the densities where the $\mu$ and Λ hyperons begin to appear, the squared speed of sound exhibits a noticeable drop. This behavior is physical, arising from the softening that occurs after a new particle emerges, although the magnitude and sharpness of the drop may depend on the specific interaction model used. However, it quickly recovers in both cases. As expected, the inclusion of hyperons (red solid line) leads to a pronounced softening of the EOS at higher densities, where the squared speed of sound remains significantly lower compared to the purely nucleonic case (blue dashed line). The hyperonic matter line tends to converge toward $c_s^2 \approx 0.6$ at higher densities, whereas the purely nucleonic case approaches $c_s^2 \approx 1$. The former reflects a more physically realistic behavior of the EOS, given that values of $c_s^2$ exceeding 1 would violate causality. It is worth noting that, in the purely nucleonic EOS, the $c_s^2$ approaches the causal limit only at densities exceeding those corresponding to the maximum stable mass of NSs. Such densities lie on the unstable branch of the mass–radius relation, where stellar configurations are no longer gravitationally stable. Consequently, the LOCVY method remains reliable throughout the entire physically relevant density range, up to the central densities of the maximum-mass configuration, without violating causality. These EOS results provide the basis for exploring their astrophysical consequences through stellar structure calculations, discussed in the next section.

## 4. Structure of Hyperonic Neutron Star

After obtaining the total energy density and pressure of the system, we solve the Tolman–Oppenheimer–Volkoff (TOV) equations for a spherically symmetric star in hydrostatic equilibrium. The mass–radius ($M$–$R$) relation is calculated for two cases: a purely nucleonic EOS and a hyperonic EOS (including Λ hyperons). The TOV equations form a coupled set of first-order differential equations as follows

$$\frac{dP(r)}{dr} = -\frac{G\rho(r)m(r)}{r^2c^2}\left(1+\frac{P}{\varepsilon(r)}\right)\left(1+\frac{4\pi r^3 P}{m(r)c^2}\right)\times\left(1-\frac{Gm(r)}{rc^2}\right)^{-1}, \tag{50}$$

$$\frac{dm(r)}{dr} = \frac{\varepsilon(r)4\pi r^2}{c^2}, \tag{51}$$

where $P(r)$, $\varepsilon(r)$, and $m(r)$ denote the pressure, the energy density, and the gravitational mass enclosed within radius $r$, respectively, and $G$ is the gravitational constant. These equations are solved simultaneously with appropriate initial conditions to obtain the stellar structure for each EOS.

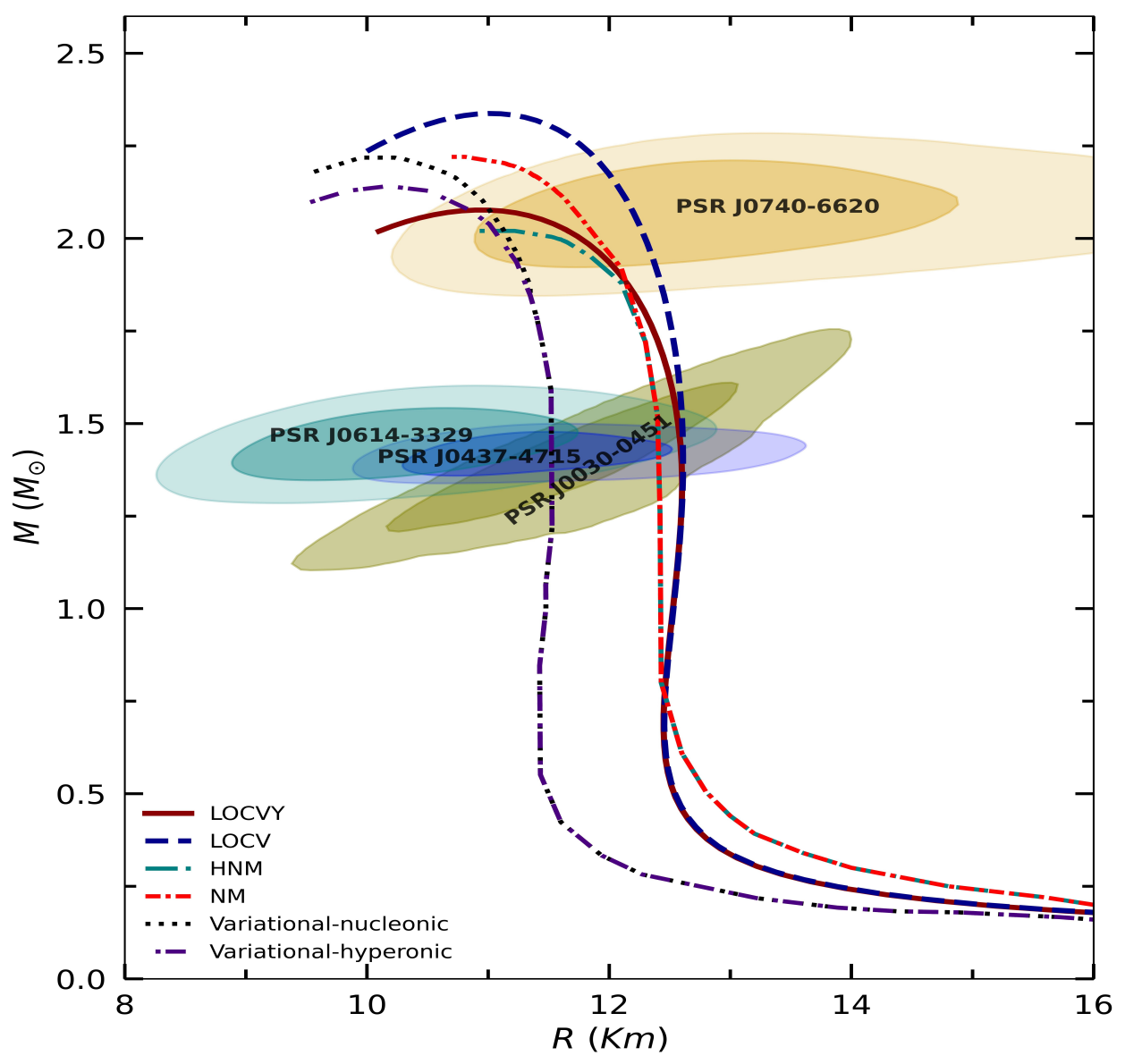

**Figure 7.** Mass–radius ($M$–$R$) relations for neutron stars with purely nucleonic cores (blue dashed line) and hyperonic cores (solid red line) within the LOCV method are presented, alongside several observational constraints. The light and dark goldenrod-shaded regions correspond to the updated $1\sigma$ and $2\sigma$ confidence intervals from the NICER analysis of PSR J0740+6620 (A. J. Dittmann et al. 2024). For PSR J0030+0451, the 95% and 68% credible intervals from NICER are shown as nested light and dark green regions (S. Vinciguerra et al. 2023). Observational constraints for PSR J0437–4715 are represented by overlapping dark and light blue regions, indicating the 68% (inner) and 90% (outer) credible levels (D. Choudhury et al. 2024a). Lastly, the most recent NICER results for PSR J0614–3329 are illustrated by nested light and dark teal regions, denoting the 95% and 68% credible intervals, respectively (L. Mauviard et al. 2025a). For comparison, we also include the $M$–$R$ relations for four additional EOSs: the HNM and the NM from H. Tong et al. (2025c), and a variational approach considering both pure nucleonic matter and hyperonic matter from H. Togashi et al. (2016).

Figure 7 shows the gravitational $M$–$R$ relation for the NSs obtained within our method for both purely nucleonic matter (blue dashed line) and hyperonic matter including $\Lambda$ hyperons (red solid line), where masses are expressed in units of the solar mass. In both cases, the DD2 model (S. Typel & S. Shlomo 2024) is used to describe the crust region of the EOS. To check the consistency of our results, the most recent observational constraints on the mass and radius of NSs obtained by the NICER (Neutron star Interior Composition ExploreR) telescope are also shown in the figure as shaded colored regions. According to Figure 7, in the purely nucleonic matter case (LOCV), the maximum stellar mass reaches $M = 2.34\,M_\odot$ at $R = 11.02$ km. When $\Lambda$ hyperons are included in the EOS (LOCVY), the maximum mass decreases to about $M = 2.07\,M_\odot$ with a corresponding radius of $R = 10.84$ km. Therefore, the main feature of the $M$–$R$ diagram is the reduction in the maximum mass of the star in the hyperonic case compared to the purely nucleonic case

Intuitively, one might expect that the inclusion of $\Lambda$ hyperons, being more massive than neutrons, would lead to an increase in the stellar mass. However, the calculations demonstrate the opposite. The discrepancy results from EOS softening caused by hyperons, which replace high-momentum nucleons and exhibit different interaction characteristics, in contrast to the stiffer behavior of the purely nucleonic EOS.

When compared with the observational constraints, both our nucleonic and hyperonic models satisfy the $1\sigma$ confidence interval from the NICER analysis of PSR J0740+6620 (E. Fonseca et al. 2021). Notably, even the hyperonic EOS reaches the lower bound of the maximum NS mass constraint, indicated by the goldenrod-shaded region. Around the canonical mass of $1.4\,M_\odot$, both the nucleonic and hyperonic curves lie within the 68% credible region for PSR J0030+0451 (S. Vinciguerra et al. 2024) and the 95% credible intervals for PSR J0437–4715 (D. Choudhury et al. 2024b) and PSR J0614–3329 (L. Mauviard et al. 2025b), all reported by the NICER mission. This distinctive feature arises from a delicate balance in the EOS obtained within the LOCVY model such that our hyperonic EOS remains sufficiently soft near $1.4\,M_\odot$ and stiff enough to support stars close to $2\,M_\odot$. Such a balance, as predicted by the extended LOCV method, helps address the hyperon puzzle.

Another notable feature of the $M$–$R$ diagram is the onset of $\Lambda$ hyperons, which occurs below $1.4\,M_\odot$, precisely where the hyperonic sequence (red solid line) begins to deviate from the nucleonic one (blue dashed line). This suggests that even low-mass NSs around $1.4\,M_\odot$ may accommodate a small fraction of hyperons within the LOCVY framework.

For comparison, Figure 7 also presents two additional $M$–$R$ curves from the literature, each based on an EOS that includes $\Lambda$ hyperons, along with their corresponding pure nucleonic versions. The HNM calculation (H. Tong et al. 2025c) based on lattice EFT and incorporating both two-body and three-body hyperonic forces, is shown as the green long dashed–dotted curve. It predicts a maximum mass over $2\,M_\odot$ and remains within the NICER $1\sigma$ confidence interval for PSR J0740+6620. Notably, when compared to its nucleonic counterpart (red short dashed–dotted curve), the $\Lambda$ onset density in HNM is higher than in our model. However, HNM yields smaller radii around $1.4\,M_\odot$ and a smaller maximum mass, indicating a softer EoS, possibly due to a more attractive hyperonic interaction. A second case, shown by the indigo dashed–dotted (black dotted) curve, corresponds to a variational approach with (without) $\Lambda$ hyperons (H. Togashi et al. 2016), incorporating hyperonic TBF. This model successfully reaches $M_{\max} > 2\,M_\odot$, but lies only marginally within the NICER $1\sigma$ constraint for PSR J0740+6620, indicating somewhat smaller radii. Notably, it falls entirely within the 68% credible intervals of all NICER observational constraints around $1.4\,M_\odot$, suggesting a softer EOS behavior compared to our model. The larger maximum mass achieved by its hyperonic EOS can be attributed to the delayed onset of $\Lambda$ hyperons, which occurs close to the $2\,M_\odot$ configuration. As a result, even the most massive stars predicted by this model contain only a small fraction of hyperons, minimizing their softening effect on the EOS. The complete numerical results for these configurations are summarized in Table 4.

Figure 8 shows the dimensionless tidal deformability as a function of the gravitational mass for both nucleonic and hyperonic NSs, as calculated using the LOCV method. The inclusion of hyperons (red solid curve) softens the EOS, leading to a systematically lower tidal deformability at a given mass, particularly for $M \gtrsim 1.4\,M_\odot$, where the $\Lambda$ hyperon fraction increases significantly. This softening reflects the reduction in pressure support due to the opening of hyperonic degrees of freedom, which increases the stellar compactness $C = GM/(Rc^2)$ and hence reduces $\Lambda \propto (1/C)^5$. Both models remain consistent with the tidal deformability constraints derived from gravitational-wave observations of the binary NS merger GW170817, detected by the LIGO and Virgo collaborations in 2017 (B. P. Abbott et al. 2018).

**Table 4**
Comparison of Maximum Mass, Radius at Maximum Mass, and Radius at 1.4 $M_\odot$ with and without Hyperons

| | $M_{\rm max}\,(M_\odot)$ | $R_{M_{\rm max}}$ (km) | $R_{1.4M_\odot}$ (km) |
|---|---|---|---|
| LOCVY | 2.07 | 10.84 | 12.46 |
| LOCV | 2.34 | 11.02 | 12.49 |
| HNM (H. Tong et al. 2025c) | 2.02 | 11.22 | 12.41 |
| NM (H. Tong et al. 2025c) | 2.22 | 10.81 | 12.41 |
| Variational-hyperonic (H. Togashi et al. 2016) | 2.14 | 10.17 | 11.50 |
| Variational-nucleonic (H. Togashi et al. 2016) | 2.21 | 10.22 | 11.50 |

The results obtained from solving the TOV equations numerically are subsequently used to plot the stellar profiles of particle fractions, pressure, and energy density. In Figure 9, the particle fraction profiles are shown for pure nucleonic matter (left panels) and hyperonic matter (right panels) for stars with masses of $M = 1.4\,M_\odot$ (top panels) and $M = 2\,M_\odot$ (bottom panels). The panels depict how the particle composition changes with radial distance from the stellar core, encompassing the entire structure from the innermost regions to the crust. In the central core (shorter distances), where densities reach their maximum, the matter may consist of purely nucleonic components or include hyperons, depending on the chosen EOS. As one moves outward, the composition changes, with heavier species disappearing as the density drops (larger distances). In the inner crust, the stellar matter becomes enriched in neutrons, eventually giving way to the outer crust dominated by electrons and fully ionized nuclei. The zoomed-in regions shown in the figures reveal that, in the outermost layers, the composition is primarily composed of ions and electrons (S. Typel & S. Shlomo 2024). At a certain radial distance within the inner crust of the star, neutron drip occurs as nuclei become unstable against neutron emission due to increasing pressure. Deeper in the star, at higher densities (or smaller distances), inverse $\beta$-decay leads to the formation of protons and electrons, establishing charge neutrality and $\beta$-equilibrium in the core. In all panels, muons are also present within the star, emerging at higher densities where the electron chemical potential exceeds the muon rest mass.

In the hyperonic matter case (right panels), $\Lambda$ hyperons appear in the core for both $M = 1.4\,M_\odot$ and $M = 2\,M_\odot$ stars, marking a key difference from the purely nucleonic case. It is noteworthy that the particle fractions of neutrons, protons, electrons, and muons remain roughly constant for $r \lesssim 10$ km, reflecting the uniform conditions in the high-density core. In this region, the EOS is very stiff, and the chemical potentials of the different species increase in a nearly parallel manner and in close coordination with other particles (see Figure 1). As a result, the $\beta$-equilibrium composition, i.e., the particle fractions, varies only weakly with radius.

The onset of $\Lambda$ hyperons occurs at different radial positions for the two masses: in the $2\,M_\odot$ star, they appear at $\approx$8.4 km (out of a total stellar radius of 11.6 km), while in the $1.4\,M_\odot$ star, they appear deeper in the core, at $\approx$4.1 km (out of a total stellar radius of 12.46 km).

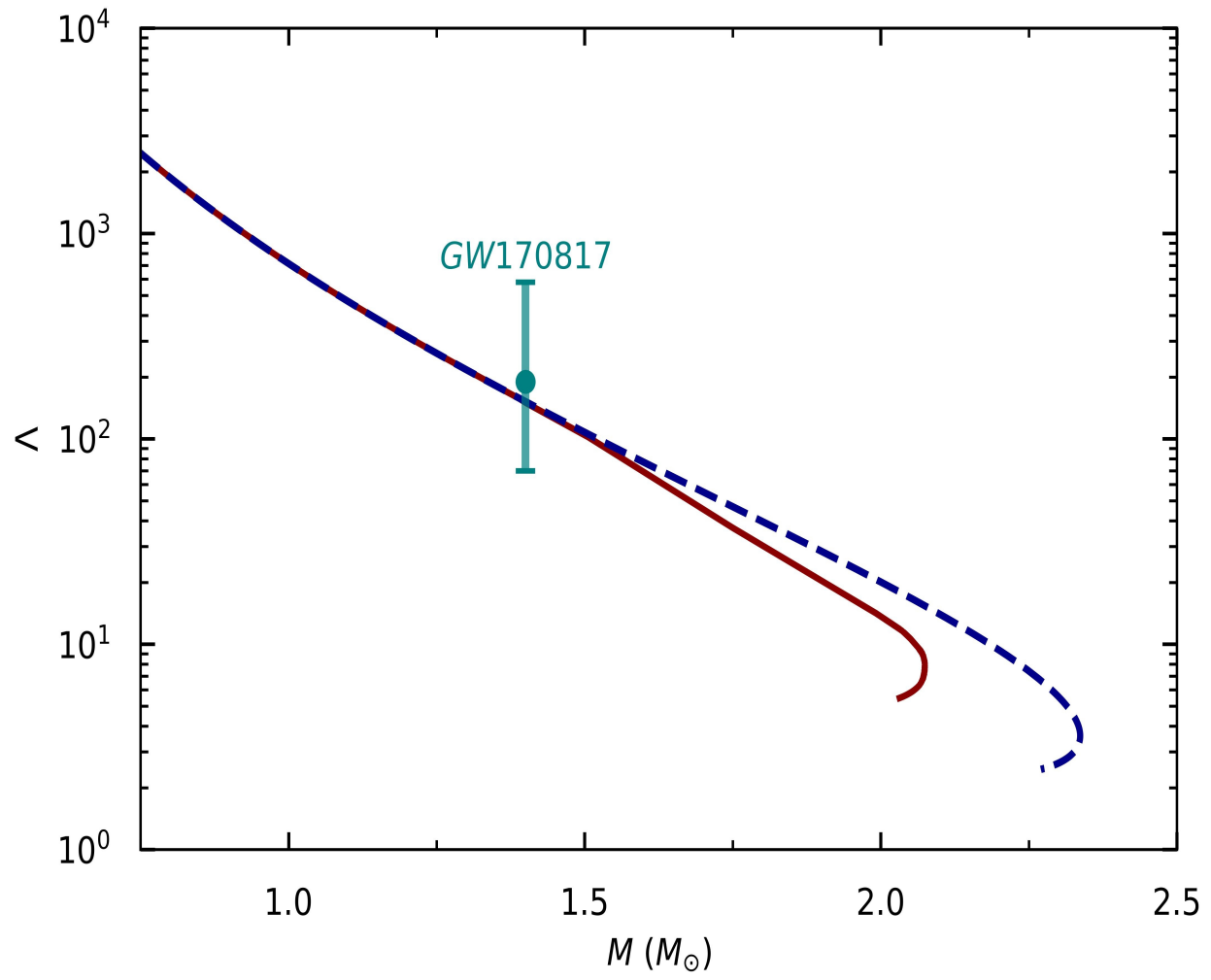

**Figure 8.** Dimensionless tidal deformability $\Lambda$ as a function of the mass for the LOCV EOSs illustrated in Figure 7. The green vertical line marks the $\Lambda_{1.4}$ constraint derived from the low-spin prior analysis of GW170817 (B. P. Abbott et al. 2018).

This difference can be attributed to the higher central density of the more massive star, so that the density crosses the $\Lambda$ threshold farther out from the center, i.e., in the $2\,M_\odot$ case, the central density is sufficiently large that the $\Lambda$ chemical potential condition is satisfied already in the outer core. In a $1.4\,M_\odot$ star, the density increases more gradually. Therefore, only the very central core reaches high enough densities to allow for hyperon formation, so the $\Lambda$ onset delays until much closer to the center.

Figure 10 shows the radial profiles of pressure (left panel) and energy density (right panel) for pure nucleonic matter (dashed lines) and hyperonic matter (solid lines). For each case, results are shown for stellar masses of $M = 2\,M_\odot$ and $M = 1.4\,M_\odot$. A clear distinction emerges between the nucleonic case and hyperonic cases. For a given stellar mass, the hyperonic EOS produces both a higher central pressure and a higher central energy density than the purely nucleonic EOS. This occurs because the hyperonic EOS is softer, making matter more compressible. To support the same stellar mass, a softer EOS requires a higher central density than a stiffer one. Consequently, hyperonic stars are more compact, with steeper pressure and energy density gradients toward the surface. These differences are directly relevant to maximum-mass constraints from recent NICER observations, which favor EOSs that can support $\sim 2\,M_\odot$ NSs while remaining consistent with radius measurements.

Another key feature evident in Figure 10 is that the primary distinction between the hyperonic and nucleonic EOSs arises in the case of the $M = 2\,M_\odot$ NS. As shown in the particle fraction profiles, the $\Lambda$ hyperon fraction remains below 0.01 at the center of the $M = 1.4\,M_\odot$ star and is therefore insufficient to noticeably affect the pressure and energy density distributions. In contrast, for the $M = 2\,M_\odot$ configuration, the $\Lambda$ fraction exceeds 0.1 and even goes beyond the proton fraction. This indicates that hyperon formation plays a much more significant role in massive stars, leading to more pronounced differences in the central pressure and energy density when compared to the purely nucleonic case.

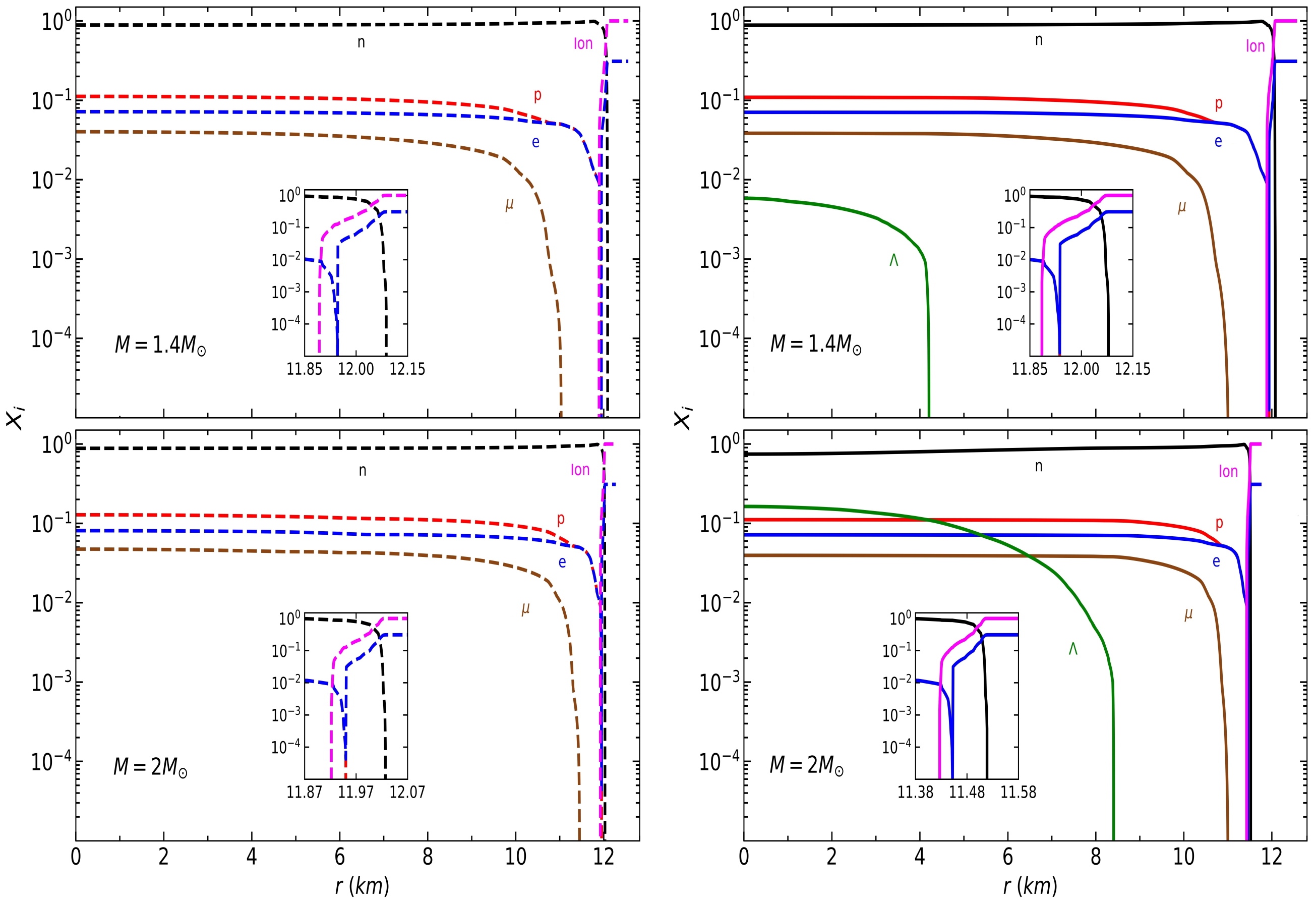

**Figure 9.** Particle fraction as a function of the distance from the stellar center. Left (right) panels correspond to nucleonic (hyperonic) matter. Top (bottom) panels show results for stars with mass $M = 1.4\,M_\odot$ ($M = 2\,M_\odot$).

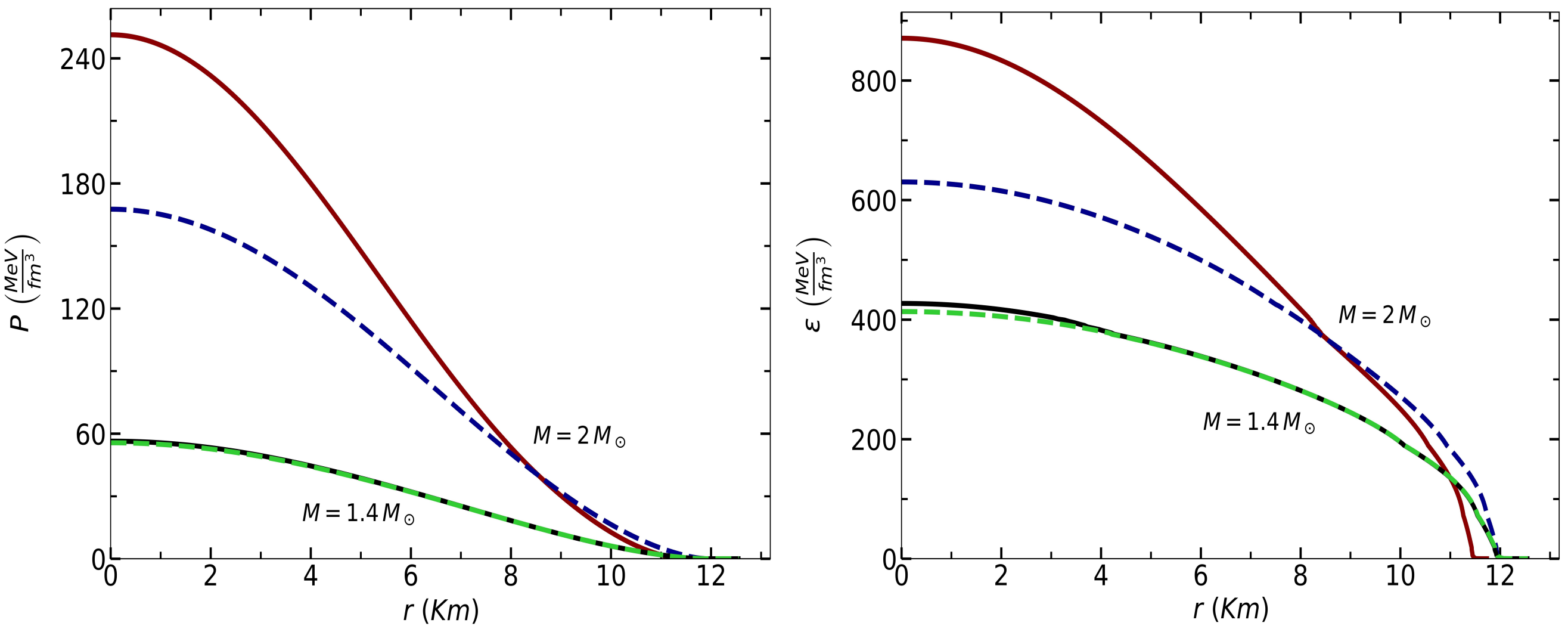

**Figure 10.** Pressure (left) and energy density (right) profiles as functions of the radius for stars with masses $M = 1.4\,M_\odot$ and $M = 2\,M_\odot$. Solid curves represent the hyperonic matter EOS, while dashed curves correspond to the pure nucleonic matter EOS.

## 5. Remarks and Conclusion

In this work, we have employed a fully microscopic framework, the LOCV method, extended to include interacting $\Lambda$ hyperons (LOCVY), to investigate the EOS and the structure of NSs. The approach is entirely based on data-constrained baryon–baryon interactions, with no adjustable parameters beyond those inherent to the two-body and TBF fitted to scattering and hypernuclear data. Our analysis results in a range of essential physical quantities, including chemical potentials, particle fractions, pressure, energy density, squared speed of sound, gravitational mass, radius, tidal deformability, and internal structure profiles of the star. Although we have restricted the hyperonic sector to the most probable species, the Λ hyperon, the resulting model remains consistent with modern observational constraints and provides valuable insight into the long-standing hyperon puzzle.

Our results indicate that Λ hyperons appear at a baryon density of about $\rho_b \approx 0.38\,\text{fm}^{-3}$, corresponding to stellar configurations below the canonical mass of $1.4\,M_\odot$. This implies that even relatively light NSs may host hyperons in their cores. The inclusion of Λ hyperons leads to a moderate softening of the EOS, reducing the maximum mass from $2.34\,M_\odot$ in the purely nucleonic case to $2.07\,M_\odot$, while still satisfying the $\sim 2\,M_\odot$ constraint from massive pulsars such as PSR J0740+6620 and the other reported high-mass candidates. The mass–radius relation obtained from our model has been evaluated considering NICER observational data for PSR J0030+0451, PSR J0437–4715, and PSR J0614–3329 to verify its consistency with the most recent astrophysical constraints. The calculated radii of the canonical mass NS and the tidal deformability are in agreement with the latest NICER measurements and with constraints derived from GW170817, further supporting the viability of the model.

From the perspective of the hyperon puzzle, these findings suggest that the appearance of Λ hyperons in NSs does not inevitably reduce the maximum mass below the observational lower bound. Within our microscopic approach, which does not yet include repulsive hyperonic TBF, hyperons can be accommodated while preserving consistency with mass, radius, and tidal deformability constraints. This outcome suggests that a careful choice of well-constrained baryon–baryon interactions, combined with a reliable microscopic treatment (such as the LOCV method), may help relieve the tension between hyperon appearance and astrophysical observations. A fully microscopic approach is particularly valuable, since it derives the EOS directly from the underlying nuclear and hypernuclear forces, minimizing reliance on phenomenological parameter adjustments.

It is important to note that the role of hyperonic TBF becomes increasingly significant at higher densities. Therefore, to achieve a more comprehensive understanding, these forces need to be incorporated into our framework to assess their impact on the EOS and the resulting mass–radius relation. In particular, reproducing single- and double-Λ hypernuclear binding energies requires the inclusion of explicit hyperonic TBF ($\Lambda NN$ and $\Lambda\Lambda N$) *as well as* a redefinition and retuning of the two-body ΛΛ interaction, whose parameters remain subject to significant uncertainties due to limited empirical constraints. Incorporating these ingredients and calibrating them against hypernuclear data is a priority for future work. In addition, other hyperons such as $\Sigma^-$ and $\Xi^-$, particularly the latter, due to its attractive single-particle potential, should be included in the formulation to evaluate their contribution to the internal structure of NSs. Taken together, our findings demonstrate that Λ hyperons can be incorporated consistently into NS matter while retaining agreement with current observational constraints, even if the hyperon puzzle is not fully resolved.

## Acknowledgments

M.S. acknowledges support from the Polish National Science Center (NCN) under grant No. 2023/48/C/ST2/00297.

## ORCID iDs

A. Mohammad Ali Looee https://orcid.org/0000-0003-0081-3151
M. Shahrbaf https://orcid.org/0000-0002-6449-8800
H. R. Moshfegh https://orcid.org/0000-0002-9657-7116